\documentstyle[12pt]{article}
\begin{document}
\newcommand{\be}{\begin{equation}}
\newcommand{\ee}{\end{equation}}
\title{Isomorphism between Non-Riemannian Gravity and Einstein-Proca-Weyl theories extended to a class of Scalar gravity theories}
\author{R. Scipioni}
\maketitle
Department of Physics, Theoretical Division, the University of Lancaster, Lancaster LA1 4YB,
England\\
and\\
Department of Physics and Astronomy, The University of British Columbia,\\
6224 Agricultural Road, Vancouver, B.C., Canada V6T 1Z1 \footnote{permanent address, e-mail: scipioni@physics.ubc.ca}
\bigskip
\bigskip
\bigskip
\begin{abstract}
We extend the recently proved relation between certain models of Non-Riemannian gravitation and Einstein-Proca-Weyl theories to a class of Scalar gravity theories, this is used to present a Black-Hole Dilaton solution with non-Riemannian connection.
\end{abstract}
PACS numbers: 04.20J, 0450, 0425.Dm, 04.40.Nr, 0470-s
\newpage
Recently much effort has been devoted to the study of non standard gravitational theories in particular some models in which both non-metricity and torsion are different from zero.\\
The Einstein's theory of gravity which was formulated more than eighty years ago provides an elegant and powerful formulation of gravitation in terms of a pseudo riemannian geometry. The Einstein's equations are obtained by considering the variation with respect to metric of the curvature scalar associated with the Levi Civita connection times the volume form of the spacetime. As assumption Einstein required that both non-metricity and torsion are vanishing, a position which is natural but not always convenient when we consider models with more degrees of freedom.\\
In particular at the level of the so called string theories [1] there are hints that by using non-Riemannian geometry we may accomodate the several degrees of freedom coming from the low energy limit of string interactions.\\
It is interesting to observe that in that case since string theories are expected to produce effects which are at least in principle testable at low energies; there may be chances to obtain non-Riemannian models with predictions which can somehow be tested, moreover some models can have some effects on astronomical scales [2].\\
For instance recently models have been proposed that permit to account for the so called dark matter by invoking short range non-riemannian gravitational interactions [3].\\
There are several approaches to non-Riemannian gravity, perhaps one of the most popular is the one which uses the gauge approach applied to the affine group [4].\\
{\bf A different approach} can be proposed in which the metric ${\bf g}$ and the connection $\bf \nabla$ are independent dynamical variables and instead of working with the affine group they rely on the definition of torsion and non-metricity in terms of $\bf g$ and $\bf \nabla$, [5].\\
Recently using this approach an interesting relation has been found between certain models of non-Riemannian gravitation and Einstein-Proca theories [6], this relation has been also found in the context of a general Metric Affine Gravity model [7].\\ 
It is the purpose of this paper to extend such relation to a class of Scalar gravity theories. This extension will be used to present a Scalar Black Hole solution with non-Riemannian connections.\\
This relation implies that though we start from a quite general non-Riemannian action the Einstein field equations simplifies into the form they would assume for a simpler action when the expression for the non-Riemannian part of the connection is used in the generalized Einstein equations.\\ 
To establish the notation which follows [8] we use a non riemannian geometry which is specified by a metric tensor field $\bf g$ and a linear connection $\bf \nabla$. using a local coframe $e^a$ with its dual frame $X_{b}$ such that $e^a(X_{b}) = \delta^a_b$, the connections 1-forms satisfy $\omega^{c}{}_{b}(X_{a}) \equiv e^c (\nabla_{X_{a}} X_{b})$. The tensor ${\bf S = \nabla g}$ defines the non metric compatibilty of the theory; in a local orthonormal frame the metric tensor is ${\bf g} = \eta_{ab} e^a \otimes e^b$, $(\eta_{ab} = diag(-1,1,1,1,...))$ The non metricity 1-forms are defined by $Q_{ab} \equiv {\bf S}(-,X_{a},X_{b})$ and the torsion 2-forms $T^a \equiv de^a + \omega^{a}{}_{b} \wedge e^b$ the curvature two forms are $R^{a}{}_{b} \equiv d\omega^{a}{}_{b} + \omega^{a}{}_{c} \wedge \omega^{c}{}_{b}$, while the general curvature scalar $R$ is given by $ R \star 1 = R^{a}{}_{b} \wedge \star (e_{a} \wedge e^b)$ in terms of the Hodge operator of the metric.\\
In ref [6] they consider the theory obtained from the action:
\begin{equation}
\Lambda[e, \omega] = k R \star 1 + \frac{\alpha}{2} dQ \wedge \star dQ + \frac{\beta}{2} Q \wedge \star Q + \frac{\gamma}{2} T \wedge \star T
\end{equation}
with k,$\alpha,\beta,\gamma$ real couplins and $Q = \eta^{ab} Q_{ab}, T = i_{c}T^{c}$ ($i_{c} \equiv i_{X_{c}}$ is the contraction operator).\\
By considering the variations induced by a coframe variation $\dot{e}^a$ and a connection variation $\dot{\omega}^{a}{}_{b}$ they obtain the field equations:
\begin{equation}
k R^{a}{}_{b} \wedge \star (e_{a} \wedge e^b \wedge e_{c}) + \tau_{c}[\alpha] + \tau_{c}[\beta] + \tau_{c}[\gamma] = 0
\end{equation}
where: 
\be
\tau_{c}[\alpha] = \frac{\alpha}{2}(dQ \wedge i_{c} \star dQ - i_{c} dQ \wedge \star dQ)
\ee
\be
\tau_{c}[\beta] = - \frac{\beta}{2}(Q \wedge i_{c} \star Q + i_{c} Q \wedge \star Q)
\ee
\be
\tau_{c}[\gamma] = \gamma [i_{k}[(T^k \wedge \star(T \wedge e_{c}))] - D i_{c} \star T - \frac{1}{2}(T \wedge i_{c} \star T + i_{c} T \wedge \star T))]
\ee
and
\be
\alpha d \star dQ + \beta \star Q = \frac{\gamma (1-n)}{2n} \star T
\ee
\be
k D \star (e_{a} \wedge e^b) = \delta^{b}{}_{a} \frac{1-n}{n} \gamma \star T + \gamma e^b \wedge i_{a} \star T
\ee
It has to be stressed that the approach we are using here of getting the field equations from variations with respect to the coframe $e^a$ and the connection $\omega^{a}{}_{b}$ is different from the usual one considered in MAG in which they consider three independent variables: $e^a, \omega^{a}{}_{b}$ and the metric $g_{ab}$ which are considered as gauge potentials in the gauging of the affine group. Since one of the equations obtained in such a way is redundant [9], it has been proposed to drop either the coframe or the metric from the list of variables [2]. We will follow that approach and choose the coframe $e^a$ and the connection $\omega^{a}{}_{b}$ as independent variables.\\
It is possible to solve equation (7) for the non-Riemannian part of the connection $\lambda^{a}{}_{b}$, this being defined by the total connection minus the levi-civita one:
\be
\lambda^{a}{}_{b} = \omega^{a}{}_{b} - \stackrel{o}{\omega}^{a}{}_{b}
\ee
If we use the solution for $\lambda^{a}{}_{b}$ in equations (2,6) it results that the field equations can be simplified into:
\be
\alpha d \star dQ + \beta_{0} \star Q = 0
\ee
and
\be
k \stackrel{o}{R}_{ab} \wedge \star(e^{a} \wedge e^b \wedge e_{c}) + \tau_{c}[\alpha] + \tau_{c}[\beta_{0}] = 0
\ee
where the superscript $o$ refers to the Levi-Civita connection and $\beta_{0}$ is:
\be
\beta_{0} = \beta - \frac{1}{4} \frac{\gamma k (n-1)^2 (n-2)}{4kn^2 (n-2) - 2 \gamma (n-1)}
\ee
So the field equations coincide with the field equations obtained from the action:
\be
\Lambda_{E-P} = k \stackrel{o}{R} \star 1 + \frac{\alpha}{2} dQ \wedge \star dQ + \frac{\beta_{0}}{2} Q \wedge \star Q 
\ee
The non metricity and torsion obtained solving equation (7) are:
\be
Q_{ab} = e_{a} i_{b} A_{1} + e_{b} i_{a} A_{1} - \frac{2}{n} \eta_{ab} A_{1} + \frac{\eta_{ab}}{n} Q
\ee
\be
T^a = \frac{1}{n-1} (e^a \wedge T)
\ee
where:
\be
A_{1} = \frac{\gamma}{nk}T
\ee
\\

We want to generalize this property of reduction to a certain class of Scalar gravity theories.\\
To begin with, consider the action:
\begin{eqnarray}
\Lambda[e,\omega] = k R \star 1 + \frac{\alpha}{2} f_{1}(\psi) (dQ \wedge \star dQ) + \\ \nonumber
\frac{\beta - \beta_{0}}{2}(Q \wedge \star Q) + f_{2}(\psi) \frac{\beta_{0}}{2} (Q \wedge \star Q) + \\ \nonumber
 \frac{\gamma}{2}(T \wedge \star T) + \frac{\delta}{2}(d \psi \wedge \star d \psi)
\end{eqnarray}
where $\psi$ is a scalar field and $f_{1}(\psi)$ and $f_{2}(\psi)$ are 0-forms functions of $\psi$.\\
We want to show that the simplification which occurs in the Einstein-Proca system occurs in the Scalar case too.\\
Considering the connection variations we get the equations:
\be
\alpha d(f_{1}(\psi) \star dQ) + f_{2}(\psi) \beta_{0} \star Q + (\beta - \beta_{0}) \star Q = \frac{\gamma(1-n)}{2n} \star T
\ee
\be
k D \star (e_{a} \wedge e^b) = \delta^{b}{}_{a} \frac{1-n}{n} \gamma \star T + \gamma e^b \wedge i_{a} \star T
\ee
since the second equation (18) is the same as (7) the same relation between $T$ and $Q$ can be obtained:
\be
T = \frac{n-1}{2n} \frac{Q}{1- \frac{2 \gamma(n-1)}{k n^2 (n-2)}}
\ee
Then provided relation (11) between $\beta, \beta_{0}, \gamma, k$ is satisfied we can write
\be
\alpha d(f_{1}(\psi) \star dQ) + f_{2}(\psi) \beta_{0} \star Q = 0
\ee
By solving (18) we can obtain the expression of the non-Riemannian part of the connection, we find that:
\begin{eqnarray}
\lambda^{a}{}_{b} = \frac{1}{2n}(e^a \wedge i_{b} [\frac{2n}{n-1}T + \frac{2 \gamma}{nk}T \\ \nonumber
- Q] + e_{b} \wedge i^{a}[Q - (2+2n)\frac{\gamma}{nk}T - \frac{2n}{n-1}T] + \delta^{a}_{b}(\frac{2 \gamma}{nk}T - Q))
\end{eqnarray}
Since the relation between $T$ and $Q$ is the same in both cases; the functional dependence of $\lambda^{a}{}_{b}$ on $Q$ will be the same, but of course the equation satisfied by $Q$ will be different, (20) instead of (9).\\
We can use the expression of $\lambda^{a}{}_{b}$ to calculate the terms in the generalized Einstein's Equations, ($G_{c} = R^{a}{}_{b} \wedge \star (e_{a} \wedge e^{b} \wedge e_{c}))$\\
\begin{eqnarray}
k {\stackrel{o}{G}}_{c}  + \\ \nonumber
f_{1}(\psi) \tau_{c}[\alpha] + f_{2}(\psi) \tau_{c}[\beta_{0}] + \tau_{c}[\delta] + \\ \nonumber
k \, \Delta G_{c}+ \tau_{c}[\beta - \beta_{0}] + \tau_{c}[\gamma] = 0
\end{eqnarray}
where $G_{c} = R^{a}{}_{b} \wedge \star (e_{a} \wedge e^{b} \wedge e_{c}))$ defines the $n-1$ Einstein forms, and:
\be
\tau_{c}[\delta] = - \frac{\delta}{2}(d \psi \wedge i_{c} \star d \psi + i_{c} d \psi \wedge \star d \psi)
\ee
Since the cancellation which occurs in the Einstein's equations depends only on the functional dependence of $\lambda^{a}{}_{b}$ on $Q$ , we can certainly say that:
\be
 k \Delta G_{c} + \tau_{c}[\beta - \beta_{0}] + \tau_{c}[\gamma] = 0
\ee
Where  $k \Delta G_{c}$ indicate the non-Riemannian correction to the Einstein $(n-1)$ forms $G_{c}$
so that the generalized Einstein's equations reduce to:
\be
k {\stackrel{o}{G}}_{c} + f_{1}(\psi) \tau_{c}[\alpha] + f_{2}(\psi) \tau_{c}[\beta_{0}] + \tau_{c}[\delta] = 0
\ee
that is, the Einstein's equations of the action:
\be
\Lambda = k \stackrel{o}{R} \star 1 + f_{1}(\psi) \frac{\alpha}{2}(dQ \wedge \star dQ) + \frac{\beta_{0}}{2} f_{2}(\psi)(Q \wedge \star Q) + \frac{\delta}{2}(d \psi \wedge \star d \psi)
\ee
Then the extension to the class of Scalar gravity theories (16) is proved.\\
To summarize we proved that though we modified action (1) by introducing factors depending on the scalar field plus a kinetic term for the dilaton field $\psi$ we obtain a simplification in the field equations (22) to (26) similar to what obtained for the equations (2) which have been simplified into (10).\\ 
We have considered the action (1) as a starting point only to keep the complexity to a reasonable level but the result is immediately generalizable to more general models like the eight parameter models considered in Ref[3] since the cancellation which occurs in the generalized Einstein's equations \emph{is not influenced by a diagonal modification of the Cartan equation} [10].\\
\\
\\
As an application of what found consider the action:
\be
\Lambda = k R \star 1 + \frac{\alpha}{2} e^{-2(\psi)}(dQ \wedge \star dQ) + \frac{\gamma}{2}(T \wedge \star T) + \frac{\delta}{2}(d \psi \wedge \star d \psi) + \frac{\beta}{2}(Q \wedge \star Q)
\ee 
The previous action is a particular case of (16) We are considering it because we want to apply what proved in the previous section to obtain a Black Hole solution using the above mentioned simplification property in the field equations. To this aim we choose $\beta_{0} = 0$ so that the Proca equation goes into the Maxwell-Dilaton equation:
\be
\alpha d(e^{-2 \psi} \star dQ) = 0
\ee
then (11) gives the constraint:
\be
4 n^2 (n-2) \beta k + (n-1)^2 (n-2) \gamma k + 8(1-n) \gamma \beta = 0
\ee
The generalized Einstein's equations become:
\be
 k \stackrel{o}{G}_{c} + \frac{\alpha}{2} e^{-2 \psi}(dQ \wedge i_{c} \star dQ - i_{c} dQ \wedge \star dQ) - \frac{\delta}{2}(d \psi \wedge i_{c} \star d \psi + i_{c} d \psi \wedge \star d \psi) = 0
\ee
the equation for $\psi$ is:
\be
\delta d \star d \psi + \alpha e^{-2 \psi} (dQ \wedge \star dQ) = 0
\ee
we consider now a spherically symmetric spacetime with metric (n=4):
\be
{\bf g} = -e^{0} \otimes e^{0} + e^{1} \otimes e^{1} + e^{2} \otimes e^{2} + e^{3} \otimes e^{3}
\ee
\begin{eqnarray}
e^{0} = f(r) dt \\ \nonumber
e^{1} = \frac{1}{f(r)} dr \\ \nonumber
e^{2} = R(r) d\theta \\ \nonumber
e^{3} = R(r) \sin \, \theta \, d\phi
\end{eqnarray}
then using the results of ref [11] we can write a solution for $Q$ and $\psi$ as:
\begin{eqnarray}
Q = - q \, cos \, \theta \, d \phi \\ \nonumber
\psi = -\frac{1}{2}ln(b_{1} - \frac{b_{2}}{r}) \\ \nonumber
R(r) = \sqrt{(r-r_{1})r} \\ \nonumber
f = \sqrt{1+ \frac{\alpha b_{1} q^2}{2 k r_{1} r}} \\ \nonumber
\delta = - 4 k \\ \nonumber
r_{1} = \frac{b_{2}}{b_{1}}
\end{eqnarray}
Solving equation (18) for $T^a$ and $Q_{ab}$ we find that the torsion is of a pure vector type $T^{a} = \frac{1}{3} e^{a} \wedge T$ and is written as:
\be
T^a = \frac{8 \beta}{9 \, \gamma}q \cos \, \theta (e^a \wedge d \phi)
\ee
the non metricity is the sum of $Q^{(3)}{}_{ab}$ a dilation piece, and $Q^{(4)}{}_{ab}$ a proper shear piece, and can be calculated using the formula:
\be
Q_{ab} = e_{a} i_{b} A_{1} + e_{b}i_{a} A_{1} - \frac{1}{2} g_{ab} A_{1} + \frac{g_{ab}}{4} Q
\ee
with:
\be
A_{1} = \frac{2 \beta q}{3 \, k} \cos \, \theta \, d\phi
\ee
\\
in conclusion if we choose the non-metricity and torsion as in (35,36) the metric (32-33), provided that (29) and (34) are satisfied, we get a Dilaton Black hole solution with non riemannian connections for the equations which come from the action (27).\\In order to have Black Hole horizon we need the condition:
\be
\frac{\alpha b_{1}}{2 k r_{1}} < 0
\ee
Then for $r = r_{0} = -\frac{\alpha b_{1} q^{2}}{2r_{1}k}$ we have an event horizon. We have to observe however that we need also another condition, namely:
\be
r_{1} \neq  - \frac{\alpha b_{1} q^2}{2k r_{1}}
\ee
otherwise for $r = r_{1}$ we would get a singular metric and the scalar field would diverge.\\
Moreover we have to observe that $Q$ is not obviously spherically symmetric but $dQ$ can be written as:
\be
dQ = q \sin \, \theta d \phi = e^{3} \frac{q}{R(r)}
\ee
so that:
\be
dQ \wedge \star dQ = \frac{q^2}{{R(r)}^2} \star 1
\ee
Using the two previous relations we can verify that although both $Q$ and $T^{a}$ are not spherically symmetric objects, equations (29,31-32) do allow for a spherical symmetric solution to exist \footnote{The situation is similar to what happens in Ref [11] where the electromagnetic potential $A= -q \cos \theta d \phi$ gives $F = q \sin \theta d \theta \wedge d \phi$ and they get spherically symmetric charged black holes}, had we chosen $\beta_{0} \neq 0$ we would have obtained equations inconsistent with the spherically symmetric ansatz. So another effect of the simplification in the field equations is the elimination of the spherically non symmetric terms which are contained in the action (27), in this respect, the condition of zero mass for the Weyl field (28) is quite relevant since it eliminates the non spherical symmetric $Q$ from the field equations (28,30,31).\\
In a paper by Matos and Macias [12] a massive Black Hole solution in dilaton gravity has been obtained starting from a 5-dimensional action in vacuum which has been reinterpreted as a four dimensional  Einstein equations with matter. This solution is considered anyway in a riemannian spacetime.\\
Our solution is somehow related to what obtained in Ref. [13-14] in the context of MAG in our case however we do not have electromagnetic field but the mathematical structure is similar since $Q$ serves as a potential to $dQ$ as $A$ for $F = dA$.\\
Recently it has been observed that by using projective invariance certain non-Riemannian models may be reduced to models containing only torsion or non-metricity [15]. Our model is not however projectively invariant, indeed a simple calculation shows that if we perform the transformation in the connection:
\be
\omega_{ab} \rightarrow \omega_{ab} + g_{ab}P
\ee
where $P$ is a constant $1$-form, 
the first two terms in (27) are invariant while the others containing the connection give:
\begin{eqnarray}
\beta (Q \wedge \star Q) + \gamma (T \wedge \star T) \rightarrow \beta (Q \wedge \star Q) + \gamma (T \wedge \star T) + \\ \nonumber
[4 \beta + \gamma (1-n)^2] (P \wedge \star P) +  P \wedge \star [4 \beta Q + 2 \gamma (1-n) T]
\end{eqnarray}
It is easy to see that even if we choose the coupling constants in a certain way it is not possible to remove the dependence on the arbitrary 1-form $P$. The consequence is that the model is not projectively invariant and in general both the torsion 1-form and the Weyl covector $Q$ are needed in our case, this must be distiguished from Eq. 6.6.3 in Ref. 4 which give nonmetricity and torsion of the almost pure gauge type in such a way either $Q$ or $T$ can be removed from the solution. Also the torsion kink solutions found by Baekler et al. [16] are physically inequivalent to ours since in there there is no nonmetricity.\\
We have to observe that recently in Poincare and Metric Affine gauge theory of gravity much effort has been devoted to the study of the problem of degeneracy of models like (1), [18-20], indeed the fact it is possible to reduce the field equations to a simpler form indicates the existence of a kind of redundancy of variables. Again it has to be observed that here we are using an approach which is different from the gauge one, so the discussion may well be completely different and will be considered in another paper.\\
Another observation regards the degrees of freedom introduced by the non-metricity (36). The expression (36) contains only two covector pieces so we avoid the causality problems [17] which may occur if we introduce a general non-Metricity with its three spacetime components, this being equivalent to the introduction a 3-spin mode.\\
There are other problems related to the models (16,27), in general they represent theories which are not renormalizable. Recently some detailed discussion has been given about the properties of the Dirac equation in Riemann-Cartan spacetime, in an interesting paper [21] it has been claimed that the chiral anomaly results to be dependent on the axial vector torsion 1-form $A = \star (e^{a} \wedge T_{a})$. In our case this 1-form vanishes identically so we may speculate that for our solutions there is no contribution from the torsion to the chiral anomaly, so the only contribution would come from the Weyl covector $Q$ besides the usual riemannian term, at least in the case in which we can \emph{adiabatically deform} the spacetime to the metric compatible case.\\
It is clear that the problem is not simple even because the Dirac equation is not well defined in a general non-Riemannian spacetime and its discussion goes beyond the aim of this paper which has been to show that certain classical non-Riemannian models appear to possess properties of reduction of the field equations which can be extended to models which contains scalar fields, this has been used  to present a black-hole solution with non-metricity and torsion.\\
Given the importance nowadays of theories containing scalar fields both in cosmology and particle physics the result seems interesting.\\
The discussion of field theoretic related problems as well as the discussion of the mathematical constraints which may origin from the requirement of a consistent cauchy formulation deserve closer inspection and will be considered in forthcoming investigations.\\ 
\\
\\
{\bf Acknowledgments}:\\
\\
This research has been partially funded by the International Center of Cultural Cooperation (NOOPOLIS), Rome ITALY.
The stimulating discussions with R. W. Tucker and C.H. Wang (Lancaster) are also acknolewdged.\\
Thanks to the referees for indicating me some of the references which were useful in revising the paper.\\
\newpage
\begin{center}
{\bf REFERENCES}
\end{center}  
1] T. Dereli, R. W. Tucker, Class. Quant. Grav. {\bf 11} (1994) 2575.\\
\\
2] R. W. Tucker, C. Wang, Class. Quant. Grav. {\bf 12} (1995) 2587.\\
\\
3] R. W. Tucker, C. Wang, Class. Quant. Grav. {\bf 15} (1998) 954.\\
\\
4] F. W. Hehl, J. D. McCrea, E. W. Mielke and Y. Ne'eman, Phys. Rep. {\bf 258} (1995) 1.\\
\\
5] C. H. Wang, PhD Thesis, Lancaster (1996).\\
\\
6] T. Dereli, M. Onder, J. Schray, R. W. Tucker and C. Wang, Class. Quant. Grav. {\bf 13} (1996) L103.\\
\\
7] Yu. N. Obhukov, E.J. Vlachysnky, W. Esser, and F. W. Hehl, Phys. Rev. D {\bf 56} 2 (1997) 7769, see also F. W. Hehl, A. Macias, gr-qc-9902076.\\
\\
8] I. Benn, R. W. Tucker, 1987 \emph{An introduction to spinors and geometry with applications in Physics} (Bristol: Hilger).\\
\\
9] F. Gronwald, Int. J. Mod. Phys. {\bf D6} (1997) 263.\\
\\
10] R. Scipioni, J. Math. Phys. (to be submitted).\\
\\
11] D. Garfinkel, G. T. Horowitz, A. Strominger, Phys. Rev. D {\bf 43} 10 (1991) 3140.\\
\\
12] T. Matos, A. Macias, Mod. Phys. Lett. A {\bf 9} (1994) 3707.\\
\\
13] J. Socorro, C. Lammerzahl, A. Macias, E. W. Mielke, Phys. Lett. A {\bf 244} (1998) 317.\\
\\
14] A. Macias, E. W. Mielke, J. Socorro, Class. Quant. Grav.{\bf 15} (1998) 445.\\
\\
15] A. Macias, E. W. Mielke, H. M.Teloti, J. Math. Phys. {\bf 36} 10 (1995) 5868.\\
\\
16] P. Baekler, E. W. Mielke, R. Hecht, F. W. Hehl, Nucl. Phys. {\bf B 288} (1987) 800.\\
\\
17] M. Kaku, \emph{Quantum Field Theory} (Oxford University Press, Oxford 1993) p.750.\\
\\
18] P. Baekler et al, Phys. Lett. {\bf 113A} (1986) 471.\\
\\
19] P. Baekler et al, Fortscr. Phys. {\bf 36} (1988) 549.\\
\\
20] R. Hecht et al, Phys. Lett. A {\bf 222} (1996) 37.\\
\\ 
21] E. W. Mielke, D. Kreimer, Int. J. Mod. Phys. D {\bf 7} 4 (1998) 535.\\
\end{document}